\documentstyle[prb,aps]{revtex}

\begin{document}
\draft
\title{Vortex structure of thin mesoscopic disks in the presence of an
inhomogeneous magnetic field}
\author{M. V. Milo\v{s}evi\'{c}, S. V. Yampolskii\cite{perm} and F. M. Peeters\cite
{aut2}}
\address{Departement \ Natuurkunde, Universiteit Antwerpen (UIA), \\
Universiteitsplein 1, \ B-2610 Antwerpen, Belgium}
\date{\today}
\maketitle

\begin{abstract}
The vortex states in a thin mesoscopic disk are investigated within the
phenomenological Ginzburg-Landau theory in the presence of different
``model'' magnetic field profiles with zero average field which may result
from a ferromagnetic disk or circulating currents in a loop near the
superconductor. We calculated the dependences of both the ground and
metastable states on the magnitude and shape of the magnetic field profile
for different values of the order parameter angular moment, i.e. the
vorticity. The regions of existence of the multi-vortex state and the giant
vortex state are found. We analysed the phase transitions between these
states and studied the contribution from ring-shaped vortices. A new
transition between different multi-vortex configurations as the ground state
is found. Furthermore, we found a vortex state consisting of a central giant
vortex surrounded by a collection of anti-vortices which are located in a
ring around this giant vortex. The limit to a disk with an infinite radius,
i.e. a film, will also be discussed. We also extended our results to
``real'' magnetic field profiles and to the case in which an external
homogeneous magnetic field is present.
\end{abstract}


\section{INTRODUCTION}

Recent progress in microfabrication and measurement techniques makes it
possible to study the properties of small superconducting structures, so
called mesoscopic samples, with sizes comparable to the penetration depth $%
\lambda $ and the coherence length $\xi $. Mesoscopic disks have been one of
the most popular and exciting study objects~\cite
{Bouisson,Leunat,Geim1,Geim2,Crit1,Schw1,Schw2,Palas1,Palas2,Akkerm,Deo,Gred,Ben}
in this respect. The behavior of such mesoscopic samples in an external
magnetic field is strongly influenced by the boundary conditions, sample
size and geometry and may lead to various superconducting states and phase
transitions between them.

Motivated by recent experiments~\cite{leuven1,leuven2}, we study the
properties of a superconducting disk in the presence of a step-like external
magnetic field. The step-like field profile is a model magnetic field
profile for a ferromagnetic dot or current loop placed on top of the
superconductor. These profiles have the important property that the average
magnetic field is zero. We investigate the influence of step height, step
profile, and ratio between step width and radius of the disk, on the
superconducting phase diagram.

Previous investigations of structures with magnetic dots were limited to
experiments with superconducting films deposited on regular arrays of
magnetic dots~\cite{leuven1,leuven2,schuller} and theoretical studies of
single magnetic dots embedded in a thin superconducting film~\cite
{Marmorkos,Fertig}. The common problem was that, for magnetic dots, the
strong field present inside the ferromagnet suppresses the superconducting
order parameter, and in such situations it is appropriate to adopt a
boundary condition in which the order parameter itself vanishes. This spoils
the effect that leads to surface superconductivity, and it is not at first
obvious why magnetic dots should support the relatively large supercurrents
associated with multiple vortices. Therefore a possible oxide layer between
the magnetic dot and the superconductor may restore the boundary condition
to the one of a superconductor/insulator interface. In the present paper we
put a single magnetic dot on top of the superconductor and study the
behavior of our sample in such non-uniform magnetic field of the dot which
enhances the possibility of obtaining various combinations of
superconducting states. To better understand the problem we start from a
simple theoretical model for the inhomogeneous magnetic field profile that,
we believe, captures all aspects of the physics involved. Models used before
vary from a representation of the magnetic dot by a perfect dipole~\cite{Hwa}
to a magnetic field profile calculated numerically for an infinitely thin
magnetic disk~\cite{Marmorkos}. To obtain a better insight we start first
with a simple step-like field model and subsequently investigate the more
complicated real magnetic field profiles, which we obtained numerically.

Theoretical studies have predicted that in mesoscopic disks surrounded by an
insulating media three kinds of superconducting states can exist - giant
vortex (a circular symmetric state with a fixed value of angular momentum),
multi-vortex state (a collection of single vortices which can be obtained as
a linear combination of giant vortices with different angular momentum), and
ring-shaped vortex states~\cite{Sergey,ring} with larger energy than giant
and multi-vortex states. The ring-shaped two-dimensional vortex states have
a cylindrically symmetric magnetic field profile and they are different from
the ring-vortices which are e.g. found in three-dimensional superfluid
liquid helium. In the present paper we observe giant vortex states and
first-order transitions between them~\cite{Geim2,Schw1}, and for
sufficiently large disks, multi-vortex structures, which are the analogue of
the Abrikosov flux-line lattice in a bulk superconductor. The latter results
not only from a mixture of giant vortex states but also from giant-ring
vortex combinations as well. The latter one may even lead to an off-center
location of a single vortex or multi-vortices. Moreover, with changing the
strength of the field there is a transition between giant-giant and
giant-ring multi-vortex states. Increasing the step height of the magnetic
field profile we found re-entrant behavior, i.e., transition from giant to
multi-vortex state and back to the giant vortex state before
superconductivity is destroyed~\cite{Schw2}. We found that for sufficiently
large magnetic disks vortex/anti-vortex structures can be formed. In order
to investigate these different vortex structures we use the method proposed
by Schweigert {\it et al.}~\cite{Schw2} and Palacios~\cite{Palas1,Palas2},
with its semi-analytical extensions of Ref.~\cite{Sergey} to determine the
stability of the different multi-vortex configurations. In particular, the
analysis of Ref.~\cite{Sergey} showed that in a superconducting disk the
ring-shaped vortices are unstable in the presence of a homogeneous magnetic
field..

Our analysis is within the framework of the phenomenological Ginzburg-Landau
(GL) theory. Although this theory has only a firm mathematical derivation in
a narrow range of magnetic field close to the superconducting-normal state
boundary, it has been found that it gives also very good results deep inside
the superconducting phase diagram~\cite{Geim2,Schw1}.

The paper is organized as follows. In Sec.~II we present our theoretical
model. In Sec.~III we discuss the giant vortex states and study the
influence of a step magnetic field profile, with zero average which is
centered at the disk or has a ring symmetry, on the superconducting state.
These step profiles are limiting cases of the actual experimentally
realizable profiles. The stability of multi-vortex states and transitions
between them are investigated in Sec.~IV. In the Appendix we give the
analytical approach/solution to this subject. The excitement in this study
is shown through different $H_{in}-R$ phase diagrams in Sec.~V, where $%
H_{in} $ is the magnitude of the magnetic field profile and $R$ represents
the radius (of superconducting disk, magnetic dot, current loop, etc.). In
Sec.~VI we present the results of our approach applied to a superconducting
disk in the presence of an experimentally realizable real magnetic dot and
current loop field profile. Sec.~VII is an extension of previous sections
where the influence of an additional homogeneous background magnetic field
is investigated. Our conclusions are given in Sec.~VIII.

\section{THEORETICAL\ MODEL}

We consider a mesoscopic superconducting disk with radius $R$ and thickness $%
d$ surrounded by vacuum. The external magnetic field $\overrightarrow{H}%
=(0,0,H)$ is directed normal to the disk plane. In this paper we investigate
two different magnetic field profiles: 1) step-like magnetic field in the
center of the disk, and 2) a ring step-like magnetic field profile with
inner radius of the ring $R_{d}$. The magnetic field strenghts of the
profile (Fig.~\ref{fig1}) are chosen such that the total magnetic flux
equals zero. These models should correspond to the magnetic field of a
perpendicular magnetized magnetic dot placed in the center of the disk on
top of the superconductor, and the magnetic field due to a current loop
placed on the superconductor, respectively. We assume that the
superconductor will not alter the magnetic state of the ferromagnetic disk.
We have to solve the system of two coupled non-linear GL equations which
determine the distribution of both the superconducting order parameter, $%
\Psi (\overrightarrow{r})$, and the magnetic field (or vector potential $%
\overrightarrow{A}(\overrightarrow{r})$) inside and outside the
superconductor 
\begin{equation}
\frac{1}{2m}\left( -i\hbar \overrightarrow{\nabla }-\frac{2e}{c}%
\overrightarrow{A}\right) ^{2}\Psi =-\alpha \Psi -\beta \Psi \left| \Psi
\right| ^{2},  \label{GL1}
\end{equation}
\begin{equation}
\overrightarrow{\nabla }\times \overrightarrow{\nabla }\times 
\overrightarrow{A}=\frac{4\pi }{c}\overrightarrow{j},  \label{GL2}
\end{equation}
where the density of the superconducting current $\overrightarrow{j}$ is
given by

\begin{equation}
\overrightarrow{j}=\frac{e\hbar }{im}\left( \Psi ^{\ast }\overrightarrow{%
\nabla }\Psi -\Psi \overrightarrow{\nabla }\Psi ^{\ast }\right) -\frac{4e^{2}%
}{mc}\left| \Psi \right| ^{2}\overrightarrow{A}.  \label{GL3}
\end{equation}
Here $\overrightarrow{r}=(\overrightarrow{\rho },z)$ \ is the
three-dimensional position in space. Due to the circular symmetry of the
disk we use cylindrical coordinates: $\rho $ is the radial distance from the
disk center, $\varphi $ is the azimuthal angle and the $z$-axis is taken
perpendicular to the disk plane, where the disk lies between $z=-d/2$ and $%
z=d/2$.

Eqs.~(\ref{GL1}-\ref{GL3}) has to be supplemented by boundary conditions
(BC) for $\Psi (\overrightarrow{r})$\ and $\overrightarrow{A}(%
\overrightarrow{r})$: 
\begin{equation}
\left. \left( -i\hbar \overrightarrow{\nabla }-\frac{2e}{c}\overrightarrow{A}%
\right) \Psi \right| _{n}=0,  \label{BC1}
\end{equation}
where the subscript $n$ denotes the component normal to the disk surface.
The boundary condition for the vector potential has to be taken far away
from the disk where the magnetic field becomes equal to the external field $%
H $ 
\begin{equation}
\left. \overrightarrow{A}\right| _{r\longrightarrow \infty }=0.
\end{equation}
Using dimensionless variables and the London gauge $div\ \overrightarrow{A}\
=0$ we can rewrite the system of equations~(\ref{GL1}-\ref{GL3}) and BC~(\ref
{BC1}) in the following form 
\begin{equation}
\left( -i\overrightarrow{\nabla }-\overrightarrow{A}\right) ^{2}\psi =\psi
-\psi \left| \psi \right| ^{2},  \label{GL1dim}
\end{equation}
\begin{equation}
-\kappa ^{2}\Delta \overrightarrow{A}=\frac{1}{2i}\left( \psi ^{\ast }%
\overrightarrow{\nabla }\psi -\psi \overrightarrow{\nabla }\psi ^{\ast
}\right) -\left| \psi \right| ^{2}\overrightarrow{A},  \label{GL2dim}
\end{equation}
\begin{equation}
\left. \left( -i\overrightarrow{\nabla }-\overrightarrow{A}\right) \psi
\right| _{n}=0.  \label{BCdim}
\end{equation}
Here all distances are measured in units of the coherence length $\xi =\hbar
/\sqrt{2m\left| \alpha \right| }$, \ the order parameter in $\Psi _{0}=\sqrt{%
\left| \alpha \right| /\beta },$ \ the vector potential in $c\hbar /2e\xi ,$ 
$\kappa =\lambda /\xi $ is \ the GL parameter, and\ $\lambda =c\sqrt{m/\pi }%
/4e\Psi _{0}$ \ is the London penetration depth. We measure the magnetic
field in $H_{c2}=c\hbar /2e\xi ^{2}=\kappa \sqrt{2}H_{c},$ where $H_{c}=%
\sqrt{4\pi \alpha ^{2}/\beta }$ is\ the thermodynamical critical field.

The free energy of the superconducting state, measured in $%
F_{0}=H_{c}^{2}V/8\pi $ units, is determined by the expression 
\begin{equation}
F=\frac{2}{V}\left\{ \int dV\left[ -\left| \psi \right| ^{2}+\frac{1}{2}%
\left| \psi \right| ^{4}+\left| -i\overrightarrow{\nabla }\psi -%
\overrightarrow{A}\psi \right| ^{2}+\kappa ^{2}\left( \overrightarrow{h}%
\left( \overrightarrow{r}\right) -\overrightarrow{H_{0}}\right) ^{2}\right]
\right\} ,  \label{F1}
\end{equation}
with the local magnetic field

\[
\overrightarrow{h}\left( \overrightarrow{r}\right) =\overrightarrow{\nabla }%
\times \overrightarrow{A}(\overrightarrow{r}). 
\]
We restrict ourselves to sufficiently thin disks such that $d\ll \xi
,\lambda .$ \ In this case, to a first approximation, the magnetic field due
to the circulating superconducting currents may be neglected and the total
magnetic field equals the external one $\overrightarrow{H_{0}}$. Within this
approximation we have to solve only the first GL equation~(\ref{GL1dim})
with $\overrightarrow{A}=\overrightarrow{A_{0}}$, where $\overrightarrow{%
H_{0}}=rot\overrightarrow{A_{0}}$. One should notice that in this approach,
for different cases, we change only the vector potential profile $%
\overrightarrow{A_{0}}$. For our step profile 
\begin{equation}
H_{0}\left( \rho \right) =\left\{ 
\begin{array}{c}
\text{ }0,\quad \quad \ \ 0\leq \rho \leq R_{d}, \\ 
H_{in},\quad \ \ \ \ R_{d}\leq \rho \leq R_{1}, \\ 
-H_{out},\text{ \ \ \ }R_{1}\leq \rho \leq R_{2}, \\ 
\text{ }0,\quad \quad \ \ R_{2}\leq \rho \leq R.
\end{array}
\right. ,
\end{equation}
the vector potential is given by 
\begin{equation}
A_{0}\left( \rho \right) =\left\{ 
\begin{array}{c}
0,\qquad \qquad \qquad \qquad \qquad \quad \quad \quad \quad 0\leq \rho \leq
R_{d}, \\ 
\frac{H_{in}}{2}\left( \rho -\frac{R_{d}^{2}}{\rho }\right) ,\qquad \qquad
\qquad \qquad \quad R_{d}\leq \rho \leq R_{1}, \\ 
\frac{H_{in}}{2\rho }\left( R_{1}^{2}-R_{d}^{2}\right) -\frac{H_{out}}{2}%
\left( \rho -\frac{R_{1}^{2}}{\rho }\right) ,\quad R_{1}\leq \rho \leq R_{2},
\\ 
0,\qquad \qquad \qquad \qquad \qquad \quad \quad \quad \quad R_{2}\leq \rho
\leq R.
\end{array}
\right. ,
\end{equation}
where $H_{in}$ describes the positive step and $%
H_{out}=-H_{in}R_{1}^{2}/(R_{2}^{2}-R_{1}^{2})$ determines the value of the
negative step (see Fig.~\ref{fig1}, Fig.~\ref{fig_lpr} and the Appendix).

First, we determine the $z$-dependence of $\psi \left( \overrightarrow{r}%
\right) $.\ Representing the order parameter as a series over cosines \ $%
\psi \left( \overrightarrow{r}\right) =\sum\limits_{k=0}^{\infty }\psi _{k}(%
\overrightarrow{\rho })\cos \left( k\pi z/d\right) $ and using the same BC~(%
\ref{BCdim}) at the disk sides $\left( z=\pm d/2\right) $ and using the
first GL equation (\ref{GL1}), one can verify that the uniform part of the
order parameter, i.e. the $k=0$ term, gives the main contribution for $(\pi
\xi /d)^{2}>>1$. Therefore, we may assume that the order parameter is
uniform along the $z$ direction of the disk and average the first GL
equation over the disk thickness. After this averaging and for fixed value
of the angular momentum it leads to $\psi (\overrightarrow{\rho })=f(\rho
)\exp (iL\varphi )$, and the problem for $f\left( \rho \right) $ is reduced
to a one dimensional problem, like in Ref.~\cite{Schw1}: 
\begin{equation}
-\frac{1}{\rho }\frac{\partial }{\partial \rho }\rho \frac{\partial f}{%
\partial \rho }+\left\langle \left( \frac{L}{\rho }-A\right)
^{2}\right\rangle f=f\left( 1-f^{2}\right) \text{,}
\end{equation}
and for the vector potential

\begin{equation}
-\kappa ^{2}\left( \frac{\partial }{\partial \rho }\frac{1}{\rho }\frac{%
\partial \rho A}{\partial \rho }+\frac{\partial ^{2}A}{\partial z^{2}}%
\right) =\left( \frac{L}{\rho }-A\right) f^{2}\theta (\rho /R)\theta
(2\left| z\right| /d)\text{,}
\end{equation}
where the function $\theta (x)=1$ ($x<1$), $0$ ($x>1$), and $R$, $d$ are the
dimensionless disk radius and thickness, respectively. The brackets $%
\left\langle ...\right\rangle $ refer to averaging over the disk thickness.

\section{GIANT\ VORTEX\ STATES}

The giant vortex state has cylindrical symmetry and consequently the order
parameter can be written as $\psi \left( \overrightarrow{\rho }\right)
=f\left( \rho \right) \exp \left( iL\varphi \right) $. The stable states are
obtained in the following way. From the {\it linearized }GL equation we find 
$f\left( \rho \right) $ up to a multiplying constant. This function is then
inserted into the free energy expression (\ref{F1}) which after minimization
determines: (i) the constant in $f\left( \rho \right) $, and (ii) the energy
value corresponding to the stable state. It can be shown that, for the case
of giant vortex states, the present approach and the one of Ref.~\cite{Schw1}%
, which was based on a solution of the non-linear GL equation, result into
the same functional $F\left( H_{in}\right) $\ dependence.

The linearized GL equation for $f\left( \rho \right) $ takes the form 
\begin{equation}
\hat{L}f=0,\quad \hat{L}=-\frac{1}{\rho }\frac{\partial }{\partial \rho }%
\rho \frac{\partial }{\partial \rho }+\left( \frac{L}{\rho }-A_{0}\right)
^{2}-1.  \label{GLlin}
\end{equation}
The superconducting state starts to develop when the minimal eigenvalue of
the operator $\hat{L}$ becomes negative. For the zero angular momentum
state, the normal state transforms to the superconducting one with
decreasing magnetic field below the nucleation field $H_{nuc}$. For nonzero
angular momentum, the superconducting state appears when we cross either the
lower $H_{nuc,l}$ or the upper $H_{nuc,u\text{ }}$ critical magnetic field
which depends on the disk radius. The eigenvalues and eigenfunctions of the $%
\hat{L}$ operator are found from 
\begin{equation}
\hat{L}f_{n,L}\left( \rho \right) =\Lambda _{n,L}f_{n,L}\left( \rho \right) 
\text{,}  \label{lin}
\end{equation}
where $f_{L,n}\left( \rho \right) $ satisfies $\left. \rho \left( \partial
f/\partial \rho \right) \right| _{\rho =0}=0$ at the disk center. The index $%
n=1,2,...$ enumerates the different states for the same $L$-value.

In general, the eigenfunctions of Eq.~(\ref{lin}) can be obtained
analytically in the case of our step magnetic field profile. We present the
complete calculation in the Appendix for both considered cases.
Alternatively, we solve Eq.~(\ref{lin}) numerically through the finite
difference technique. We put the order parameter on a space grid and find
numerically the eigenfunctions and eigenvalues of the operator $\hat{L}$
using the Householder technique.

We start our analysis with the magnetic field profile shown in Fig.~\ref
{fig1}. First, we consider a magnetic dot on top of the center of the disk,
i.e. with $R_{d}=0$, and $R_{2}=R=6.0\xi $ and study the influence of $R_{1}$%
, i.e. the width of the positive field region, on the superconducting state.

The magnetic field dependence of $\Lambda $ for different angular momenta $L$
are shown in Figs.~\ref{fig2}(a-c) for the lowest radial state, i.e. $n=0$,
and in Figs.~\ref{fig2}(d-f) for the first radial state, i.e. $n=1$ for
three different values of $R_{1}$. The top axis shows the flux corresponding
to the positive magnetic field region $\phi _{in}=H_{in}\pi R_{1}^{2}$,
which in dimensionless units becomes $\phi _{in}/\phi
_{0}=(H_{in}/2H_{c2})(R_{1}/\xi )^{2}$ where $\phi _{0}=ch/2e$ is the
quantum of flux. All numerical calculations were done for a disk thickness $%
d/\xi =0.1$ which is within the thin disk approximation. The negative $L$
values (dashed curves in Fig.~\ref{fig2}) correspond to ``anti-vortices'' in
conventional superconductors. The thin horizontal line gives the $\Lambda =0$
level. From Fig.~\ref{fig2} one notices that with increasing $R_{1}$: 1)~the
eigenvalues $\Lambda $ of the states with the same $L$ become more negative;
2)~the magnetic field range over which solutions of Eq.~(\ref{GLlin}) can be
found decreases; 3)~the number of possible solutions decreases also; and 4)
the $\Lambda (H_{in})$ dependences become more parabolic. The latter can be
explained by the fact that increasing $R_{1}$ corresponds to a more
homogeneous magnetic field profile inside the disk, i.e. we reach the case
considered in Refs.~\cite{Schw1,Sergey}.

For small $R_{1}$ the curves $L$ and $L+1$ anti-cross for sufficiently large 
$L$ values and consequently the low vorticity states have lower energy even
with increasing strength of the magnetic field profile. E.g. for $R_{1}/\xi
=1.5$ this occurs when $L>4=L^{\ast }$ and for $R_{1}/\xi =3.0$ when $%
L>5=L^{\ast }$. Notice that the slope of $\Lambda $ for the $L>L^{\ast }$
curves is substantially smaller than for the $L\leq L^{\ast }$ curves in the
high field region. Notice also that the $n=1$ states have a higher $\Lambda $
value than the $n=0$ states and their energy is also larger than those of
the anti-vortex states for the same vorticity.

In Figs.~\ref{fig_psi}(a-c) the radial dependence of the superconducting
density $\left| \psi \right| ^{2}$ is shown for $\left| L\right| <3$ at $%
H_{in}=0.75H_{c2}$ for the corresponding profiles of Fig.~\ref{fig2}. Notice
that, with increasing $R_{1}$, the Cooper-pair density near the edge of the
sample becomes more non-homogeneous. The $\left| L\right| \neq 0$ states
have a vortex sitting in the center of the disk, i.e. $\psi (\rho =0)=0$,
which becomes larger with increasing $\left| L\right| $. For the situation
of Fig.~\ref{fig_psi}(c) there is a narrow, very negative magnetic field
region of $H_{out}/H_{c2}=0.9643$ in the region $4.5<\rho /\xi <6.0$ which
leads to a considerable suppression of the Cooper pair density. For example,
the $L=0$ vortex state, i.e. the Meissner state, has a strongly reduced
Cooper pair density for $\rho /\xi >1$. For the excited state $\left(
n,L\right) =\left( 1,0\right) $ (dashed curves in Fig.~\ref{fig_psi})\ the
order parameter vanishes inside the disk and a ring-shaped node in the wave
function $\psi $ is formed~\cite{Sergey,ring} which leads to a ring-like
vortex.

The eigenvalues $\Lambda $ determine the free energy $F$ of the giant vortex
state. For the giant vortex state we consider only states which lie below
the $F=0$ level. In this approximation the order parameter is 
\begin{equation}
\psi \left( \overrightarrow{\rho }\right) =\left( -\Lambda \frac{I_{2}}{I_{1}%
}\right) ^{1/2}f_{n,L}\left( \rho \right) \exp \left( iL\varphi \right) 
\text{,}
\end{equation}
and the minimal energy value is 
\begin{equation}
F=-\Lambda ^{2}\frac{2\pi d}{V}\frac{I_{2}^{2}}{I_{1}}\text{,}  \label{Fmin}
\end{equation}
where 
\[
I_{1}=\int\limits_{0}^{R}\rho d\rho \ f_{n,L}^{4}\left( \rho \right) ,\quad
I_{2}=\int\limits_{0}^{R}\rho d\rho \ f_{n,L}^{2}\left( \rho \right) \text{.}
\]

The dependence of the free energy on the magnetic field strength of the
inner core of the magnetic field profile, $H_{in}$, are shown in Figs.~\ref
{Fig3}(a-c) for the angular momenta $\left| L\right| <11$ for different
values of $R_{1}$. The free energy is expressed in units of $%
F_{0}=H_{C}^{2}V/8\pi $. The highest value of vorticity in this disk is $%
L=15 $ (see Figs.~\ref{fig2}(a-c)). From a comparison of the magnetic field
dependence of $F\left( H_{in}\right) $ for $R_{1}/\xi =1.5$ (Fig.~\ref{Fig3}%
(a)) and the one with $R_{1}/\xi =4.5$ (Fig.~\ref{Fig3}(c)) we clearly
observe the reduction of superconductivity and the reduction of the maximal
possible vorticity with increasing $R_{1}$. The envelope of \ the lowest
parts of the curves in Fig.~\ref{Fig3} represents the field dependence of
the ground state energy. Notice that the increase of the width of the
positive magnetic field region leads to an increase of the energy of the
ground state. With increasing applied field the $L\longrightarrow L+1$
vortex transitions take place at the field where the corresponding curves
cross (for example, the $0\longrightarrow 1$ transition occurs at \ $%
H_{in}=0.8705H_{c2}$ for $R_{1}/\xi =1.5$). The crossing points are shifted
towards lower field values when increasing $R_{1}$. The $L\longrightarrow
L+1 $ transitions are of first order and lead to jumps in the magnetization
of the sample. Notice that the positive flux captured in the superconducting
disk for different $L$-states is not quantized. This is made very clear in
the $\phi _{in}-R_{1}$ diagram presented in Fig.~\ref{Fig3}(d) which shows
the ground state vortex configurations. Notice that the flux in the positive
magnetic field region has to increase with more than one flux quantum $\phi
_{0}$ before the vorticity of the superconducting state increases with one
unit. A similar phenomena was observed earlier for mesoscopic disks and
rings in a homogeneous magnetic field~\cite{ringben}. Notice also that for
higher $L$ values quantization is slowly restored, and $\Delta \phi $
decreases with enlarging $R_{1}$, i.e. the radius of the positive field
region.

Following our assumption that field inside the superconductor equals the
external one, we obtain the expression for the superconducting current
density~\cite{Sergey}: 
\begin{equation}
j_{n,L}=\frac{\Lambda _{n,L}B_{n,L}}{A_{n,L}}(\frac{L}{\rho }%
-A_{0})f_{n,L}^{2}(\rho )
\end{equation}
where $\Lambda _{n,L}$ is determined by Eq.~(\ref{lin}), $A_{0}$ represents
the vector potential of the applied field and

\[
A_{n,L}=\frac{2\pi d}{V}\int\limits_{0}^{R}\rho d\rho \ f_{n,L}^{4}\left(
\rho \right) \text{, }B_{n,L}=\frac{2\pi d}{V}\int\limits_{0}^{R}\rho d\rho
\ f_{n,L}^{2}\left( \rho \right) \text{.} 
\]
The magnetic field due to the supercurrents, neglected in our first order
approximation, is calculated from 
\begin{equation}
rot\overrightarrow{H_{sc}}=\frac{1}{\kappa ^{2}}\overrightarrow{j}\text{.}
\end{equation}
Since the supercurrent has only an azimuthal component, and, is situated
only in the superconductor plane, the $\rho $ component of $\overrightarrow{%
H_{sc}}$ can be neglected. Consequently, we obtain 
\begin{equation}
H_{sc}(\rho )=-\frac{1}{\kappa ^{2}}\int j_{\varphi }(\rho )d\rho \text{,}
\end{equation}
and the magnetization of the superconductor is then defined as the magnetic
field expelled from the superconductor. 
\begin{equation}
M=\int \frac{H_{total}-H_{0}}{8\pi }dV=\frac{d}{4}\int\limits_{0}^{R}H_{sc}(%
\rho )\rho d\rho \text{,}
\end{equation}
where $H_{0}$ denotes the applied magnetic field, and $d$ is the disk
thickness.

The corresponding $M\left( H_{in}\right) $ curves are given as insets in
Figs.~\ref{Fig3}(a-c). The phase transition from the superconducting to the
normal state is of second order (all curves $F\left( H_{in}\right) $ reach
the $F=0$ line with zero derivative). The curves $F\left( H_{in}\right) $ in
Fig.~\ref{Fig3} which are situated above the ground state energy correspond
to metastable giant vortex states. With increasing applied field the
transition from the Meissner state ($L=0$) to the normal state goes through
a set of consecutive first order transitions between the $L$ and $L+1$ giant
vortices which is finished by a second order transition to the normal state.

Our next step was to fix the magnetic field profile (we took $R_{d}/\xi =0.0$%
, $R_{1}/\xi =4.5$, $R_{2}/\xi =6.0$) and enlarge the disk radius $R$. In
Figs.~\ref{fig_psi1}(a-b) the radial dependence of the superconducting
density $\left| \psi \right| ^{2}$ is shown for $\left| L\right| <3$ at $%
H_{in}=0.75H_{c2}$ and for a superconducting disk of radius $R/\xi =9.0$,
and $12.0$, respectively. One can see that, even for $L=0$,
superconductivity in the center of the disk is destroyed, which is opposite
to the homogeneous magnetic field case where $\left| \Psi \right| ^{2}$ is
maximal at $\rho =0$~\cite{Schw1}. The shape of the $L=0$ curve changes
drastically and this occurs already for small enlargement of the disk. This
transition is shown in Fig.~\ref{fig_psi2}(a). By increasing $R/\xi $ from
6.1 to 6.5 the Cooper pair density in the center of the superconducting disk
decreases from 0.73 to 0.03. Furthermore, in Fig.~\ref{fig_psi1}(b) we see
also qualitative changes in the Cooper pair density of the excited state $%
\left( n,L\right) =\left( 1,0\right) $ (dashed curve). Now the order
parameter vanishes twice inside the disk and a double ring-like vortex is
formed. This transition is shown in Fig.~\ref{fig_psi2}(b) with increasing
disk size from $R/\xi =10.0$ to $11.0$.

The dependences of the free energy on the magnetic field $H_{in}$ for
different sizes of the superconducting disk are shown in Figs.~\ref{Fig5}%
(a-d). Notice that there are large differences in comparison with the
previous cases. We still have the $L\longrightarrow L+1$ transitions, which
are of first order and which lead to jumps in the magnetization of the
sample (see Figs.~\ref{Fig6}(a-c)), but with increasing applied field the
transition from the Meissner state ($L=0$) to the normal state goes through
a set of consecutive first order transitions and is finished by a first
order transition back to the $L=0$ state, sometimes with an intermediate $%
L=1 $ state. For $R/\xi =9.0$ we have $0\rightarrow 1\rightarrow
2\rightarrow 3\rightarrow 1\rightarrow 0$ transitions, for $R/\xi =12.0$ we
have $0\rightarrow 1\rightarrow 2\rightarrow 1\rightarrow 0$, and for $R/\xi
=18.0$ only $0\rightarrow 1\rightarrow 0$. Moreover, for a sufficiently
large disk (see Fig.~\ref{Fig5}(d) which corresponds practically to the $%
R/\xi \rightarrow \infty $ situation) we have no first order transitions but
only the Meissner state as the ground state. The reason is that for $R/\xi
\rightarrow \infty $ the non-zero magnetic field region is limited
(relatively) to a small area and we can always define a circle with radius $%
\rho $ sufficiently large where the superconducting current is zero and,
consequently one must have $L=0$. But, this Meissner state is qualitatively
different from the ''usual'' one. The radial distribution of the Cooper pair
density is extremely inhomogeneous and superconductivity is strongly
suppressed in the interior of the disk (see Figs.~\ref{fig_psi1}(b) and \ref
{fig_psi2}(a)).

\section{MULTI-VORTEX\ STATES}

It is well known that for sufficiently large disks the giant vortex state
can break up into multi-vortices~\cite{Schw2,Palas1,Palas2}. In our
analysis, we considered different field profiles and disk geometries to
investigate the properties of this transition. For smaller disks, the
confinement effect dominates and we found that only the giant vortex states
are stable and possible multi-vortex states, if they exist, have always
larger energies. In order to investigate such structures in our case we use
the method proposed by Schweigert {\it et al}.~\cite{Schw2} and Palacios~ 
\cite{Palas1,Palas2} and extended it to determine the stability of the
different multi-vortex configurations as proposed in Ref.~\cite{Sergey}.
Following Refs.~\cite{Schw2,Palas1,Schw3} the order parameter of the
multi-vortex state is written as a linear combination of eigenfunctions of
the linearized GL equation~(\ref{GLlin}) 
\begin{equation}
\psi \left( \overrightarrow{\rho }\right)
=\sum\limits_{L_{j}=0}^{L}\sum\limits_{n=0}^{\infty
}C_{n,L_{j}}f_{n,L_{j}}\left( \rho \right) \exp \left( iL_{j}\varphi \right)
,  \label{SP}
\end{equation}
where $L$ is, in the homogeneous magnetic field case, the value of the
effective total angular momentum which is equal to the number of vortices in
the disk, and $n$ enumerates the different radial states for the same $L_{j}$%
. Later on, we will see that for our inhomogeneous magnetic field case the
assignment of the total vorticity can be more tricky.

Substituting (\ref{SP}) in the free energy expression (\ref{F1}) we obtain $%
F $ as a function of the complex parameters \{$C_{n,L_{j}}$\}. Minimization
of $F$ with respect to these parameters allows us, to find the equilibrium
vortex configurations, and to determine their stability. We use the
procedure described in Refs.~\cite{Sergey,Schw3} with full consideration of
combinations of vortex configurations with different radial states.

We begin our analysis with states built up by only two components in Eq.~(%
\ref{SP}). This brings quantitative bounds in our calculation but,
nevertheless, will give the correct qualitative behavior and facilitates the
physical insight into the problem. The free energy of a two component state
built out of $(n_{1},L_{1})$ and $(n_{2},L_{2})$ is 
\begin{equation}
F=C_{n_{1},L_{1}}^{4}A_{n_{1},L_{1}}+C_{n_{2},L_{2}}^{4}A_{n_{2},L_{2}}+4C_{n_{1},L_{1}}^{2}C_{n_{2},L_{2}}^{2}A_{n_{1},n_{2},L_{1},L_{2}}+2\Lambda _{n_{1},L_{1}}C_{n_{1},L_{1}}^{2}B_{n_{1},L_{1}}+2\Lambda _{n_{2},L_{2}}C_{n_{2},L_{2}}^{2}B_{n_{2},L_{2}},
\label{F2}
\end{equation}
where 
\begin{eqnarray*}
A_{n_{i},L_{i}} &=&\frac{2\pi d}{V}\int\limits_{0}^{R}\rho d\rho \
f_{n_{i},L_{i}}^{4}\left( \rho \right) , \\
A_{n_{1},n_{2},L_{1},L_{2}} &=&\frac{2\pi d}{V}\int\limits_{0}^{R}\rho d\rho
\ f_{n_{1},L_{1}}^{2}\left( \rho \right) f_{n_{2},L_{2}}^{2}\left( \rho
\right) , \\
B_{n_{i},L_{i}} &=&\frac{2\pi d}{V}\int\limits_{0}^{R}\rho d\rho \
f_{n_{i},L_{i}}^{2}\left( \rho \right) \text{.}
\end{eqnarray*}
One should notice that we leave the possibility of combination of states
with different radial states, i.e. $n_{1}$ and $n_{2}$. Although, in
general, $C_{n_{i},L_{i}}$ is a complex number, for our two component state $%
C_{n_{i},L_{i}}$ is a real number. Minimization of Eq.~(\ref{F2}) with
respect to $C_{n_{1},L_{1}}$ and $C_{n_{2},L_{2}}$\ gives for the
multi-vortex states: 
\begin{eqnarray}
C_{n_{1},L_{1}}^{\left( 0\right) } &=&\pm \left( \frac{-\Lambda
_{n_{1},L_{1}}A_{n_{2},L_{2}}B_{n_{1},L_{1}}+2\Lambda
_{n_{2},L_{2}}A_{n_{1},n_{2},L_{1},L_{2}}B_{n_{2},L_{2}}}{%
A_{n_{1},L_{1}}A_{n_{2},L_{2}}-4A_{n_{1},n_{2},L_{1},L_{2}}^{2}}\right)
^{1/2},  \label{sol3} \\
C_{n_{2},L_{2}}^{\left( 0\right) } &=&\pm \left( \frac{-\Lambda
_{n_{2},L_{2}}A_{L_{1},n_{1}}B_{n_{2},L_{2}}+2\Lambda
_{L_{1},n_{1}}A_{n_{1},n_{2},L_{1},L_{2}}B_{L_{1},n_{1}}}{%
A_{n_{1},L_{1}}A_{n_{2},L_{2}}-4A_{n_{1},n_{2},L_{1},L_{2}}^{2}}\right)
^{1/2},  \nonumber
\end{eqnarray}
and inserting these expressions into Eq.~(\ref{F2}) leads to the energy of
the {\it multi-vortex state} 
\begin{equation}
F_{n_{1},n_{2},L_{1},L_{2}}=\frac{-\Lambda
_{n_{1},L_{1}}^{2}A_{n_{2},L_{2}}B_{n_{1},L_{1}}^{2}-\Lambda
_{n_{2},L_{2}}^{2}A_{n_{1},L_{1}}B_{n_{2},L_{2}}^{2}+4\Lambda
_{n_{1},L_{1}}\Lambda
_{n_{2},L_{2}}A_{n_{1},n_{2},L_{1},L_{2}}B_{n_{1},L_{1}}B_{n_{2},L_{2}}}{%
A_{n_{1},L_{1}}A_{n_{2},L_{2}}-4A_{n_{1},n_{2},L_{1},L_{2}}^{2}}
\end{equation}
The corresponding conditions for the stability of the vortex state are 
\begin{eqnarray}
\frac{\partial ^{2}F}{\partial C_{n_{1},L_{1}}^{2}} &=&\frac{%
8A_{n_{1},L_{1}}\left( -\Lambda
_{n_{1},L_{1}}A_{n_{2},L_{2}}B_{n_{1},L_{1}}+2\Lambda
_{n_{2},L_{2}}A_{n_{1},n_{2},L_{1},L_{2}}B_{n_{2},L_{2}}\right) }{%
A_{n_{1},L_{1}}A_{n_{2},L_{2}}-4A_{n_{1},n_{2},L_{1},L_{2}}^{2}}>0, \\
\frac{\partial ^{2}F}{\partial C_{n_{2},L_{2}}^{2}} &=&\frac{%
8A_{n_{2},L_{2}}\left( -\Lambda
_{n_{2},L_{2}}A_{n_{1},L_{1}}B_{n_{2},L_{2}}+2\Lambda
_{n_{1},L_{1}}A_{n_{1},n_{2},L_{1},L_{2}}B_{n_{1},L_{1}}\right) }{%
A_{n_{1},L_{1}}A_{n_{2},L_{2}}-4A_{n_{1},n_{2},L_{1},L_{2}}^{2}}>0, 
\nonumber
\end{eqnarray}
\begin{eqnarray*}
&&\frac{\partial ^{2}F}{\partial C_{n_{1},L_{1}}^{2}}\cdot \frac{\partial
^{2}F}{\partial C_{n_{2},L_{2}}^{2}}-\left( \frac{\partial ^{2}F}{\partial
C_{n_{1},L_{1}}\partial C_{n_{2},L_{2}}}\right) ^{2} \\
&=&\frac{64\left( -\Lambda
_{n_{1},L_{1}}A_{n_{2},L_{2}}B_{n_{1},L_{1}}+2\Lambda
_{n_{2},L_{2}}A_{n_{1},n_{2},L_{1},L_{2}}B_{n_{2},L_{2}}\right) }{%
A_{n_{1},L_{1}}A_{n_{2},L_{2}}-4A_{n_{1},n_{2},L_{1},L_{2}}^{2}}\left(
-\Lambda _{n_{2},L_{2}}A_{n_{1},L_{1}}B_{n_{2},L_{2}}+2\Lambda
_{n_{1},L_{1}}A_{n_{1},n_{2},L_{1},L_{2}}B_{n_{1},L_{1}}\right) >0.
\end{eqnarray*}

In our analysis we investigated the influence of the width of the positive
field region and radius of the disk on the phase diagram, and especially on
the stability of the multi-vortex states. For small values of $R_{1}/\xi $
these multi-vortex states are always metastable. However, with enlarging $%
R_{1}/\xi $ these states can obtain lower energy. The energies of the
equilibrium vortex states are plotted in Fig.~\ref{Fig8}(a) for $R_{1}/\xi
=5.25$ and $R_{2}/\xi =R/\xi =6.0$ and in Fig.~\ref{Fig8}(b) for $R_{1}/\xi
=4.5$ and $R_{2}/\xi =6.0$ with a larger disk radius $R/\xi =9.0$. The giant
vortex states are given by solid curves, anti-vortex states by dashed and
the multi-vortex states (dotted curves) by $\left( L_{1}:L_{2}\right) $,
i.e. the angular momentum values they are composed of. As expected, when
enlarging the disk, multi-vortex states become more stable and, moreover in
Fig.~\ref{Fig8}(b), we observe the existence of multi-vortices as a
combination of a giant and a ring-like vortex. These states have
surprisingly low energy and become the ground state for specific ranges of
magnetic field. Furthermore, when increasing the field, we observe a phase
transition between this giant-ring and giant-giant multi-vortex states, when
another type of multi-vortices becomes the ground state. The giant-ring
multi-vortex states with lowest energy are given in Fig.~\ref{Fig8}(b) by
the quantum numbers $\left( (n_{1},L_{1}):\left( n_{2},L_{2}\right) \right) $%
. It should be noted that there are many other metastable combinations
possible, which are not shown in the figures.

As shown in Fig.~\ref{Fig8}, with increasing $R_{1}/\xi $ and the size of
the superconducting disk the multi-vortices become more stable and they can
even become the minimum of the $F\left(
C_{n_{1},L_{1}},C_{n_{2},L_{2}}\right) $ function. We focus our attention to
Fig.~\ref{Fig8}(a). The solid and dashed curves represent the giant vortex
and anti-vortex states, respectively. The dotted curves are the energies of
the multi-vortices. For example, lets follow the $L=0$ curve which splits
into the $(0:4)$ multi-vortex state, at $H_{in}/H_{c2}=0.28$ and with
further increase of the field becomes another giant vortex state, but now
with $L=4$, at $H_{in}/H_{c2}=1.0$. The concomitant change of the Cooper
pair density is illustrated in Fig.$~$\ref{Fig_ent}. Vortices enter the disk
from the boundary and move to the middle with increasing field, and join
into a giant vortex again.

For a fixed magnetic field profile and with increasing size of the
superconducting disk, we obtain a variety of different superconducting
states. Let us discuss the giant-ring configurations first. We observe
transitions between the giant and multi-vortex states with the {\it same}
vorticity. The contour plots of the $\left| \psi \right| ^{2}$ distribution
for different multi-vortex states as ground states ($\left( \left(
0,1\right) :(1,0)\right) $, $\left( \left( 0,2\right) :(1,0)\right) $, $%
\left( \left( 0,3\right) :(1,0)\right) $) are shown in Figs.~\ref{Fig9}(a-c)
for a magnetic field profile with $R_{1}/\xi =4.5$, $R_{2}/\xi =6.0$ and $%
R/\xi =9.0$, for different values of $H_{in}$ (dark regions correspond to
high density and white regions to low density). The most interesting result
we obtain for the $\left( \left( 0,1\right) :(1,0)\right) $ combination.
This configuration becomes the ground state configuration for $%
H_{in}/H_{c2}=0.75$. Surprisingly, the contour plot of the $\left| \psi
\right| ^{2}$ distribution is not circular symmetric - there is one vortex
positioned off-center. In the lower part of the same figures we give the
corresponding contour plots of the phase of the superconducting wave
functions. The contour plot of the phase also nicely illustrates the
off-center location of the vortex. For the two other cases more vortices are
present which are located on a ring centered around the center of the
superconducting disk. The giant-ring multi-vortex state can be seen as a
transition between $\left( 1,0\right) $ and $\left( 0,5\right) $ state (see
Fig.~\ref{Fig8}(b)). As function of the magnetic field we start from a ring
vortex, for $H_{in}/H_{c2}=0.03$, this state splits into a giant-ring vortex
state, and finally, for $H_{in}/H_{c2}=0.98$ we obtain a giant vortex state.
This remarkable phenomenon is illustrated in Fig.~\ref{Fig11}.

However, giant-giant vortex combinations show completely different
behavior.\ As shown in Fig.~\ref{Fig8}(b) with increasing magnetic field $%
H_{in}$, combinations $\left( 0:7\right) $, $\left( 0:8\right) $, $\left(
0:9\right) $ etc. become the ground state. In the contour plot of the $%
\left| \psi \right| ^{2}$ distribution we observe single vortices which are
arranged on a ring with a low density area situated in the center of the
disk (see Figs.~\ref{Fig12}(a-c)). This central area is associated with a
giant vortex, and encircling it leads to a phase change showing the
vorticity $7$. Fig.~\ref{Fig12}(d) shows the contour plot of the
corresponding phase with seven {\it anti-vortices} arranged in a circle
around the giant vortex and the {\it total} vorticity is $L=0$. Figs.~\ref
{Fig12}(e) and Fig.~\ref{Fig12}(f) correspond, respectively, to a giant
vortex state with vorticity $8$ and $9$ where, respectively, $8$ and $9$
anti-vortices are located between the giant vortex and the boundary. Notice
that the Cooper pair density for the $L=0$ state for this disk geometry ($%
R_{1}/\xi =4.5$, $R_{2}/\xi =6.0$, $R/\xi =9.0$) shows a similar behavior
with a low density area in the center of the disk (see Fig.~\ref{fig_psi1}%
(b)). It is obvious that enlarging the superconducting disk enhances the
influence of the negative part of the step-like magnetic field. In the case
when multi anti-vortices are involved, the {\it total} vorticity of the
giant-giant multi-vortex state is equal to the lowest vorticity of the two
giant vortex states which compose the vortex state.

One more result should be noted. Following the ground state free energy
diagram (Fig.~\ref{Fig8}(b)) we notice that the total vorticity doesn't
change uniformly ($0\rightarrow 1\rightarrow 2\rightarrow 3\rightarrow
1\rightarrow 7\rightarrow 8\rightarrow 9$) and that the jumps in vorticity $%
\left| \Delta L\right| $ are not always equal to $1$. This differs from
Ref.~ \cite{Palac3} where it was claimed that the lowest barriers are those
between the $L$ and $L\pm 1$ states. We found that the $L\longrightarrow
L\pm 1$ transitions take place in disks for a small maximum value of
vorticity or at magnetic fields close to the ``superconducting-normal
state'' transition point. Between these two limiting regimes $%
L\longrightarrow L\pm N$ transitions are possible with $\left| \Delta
L\right| =N>1$. Our results are also in agreement with Ref.~\cite{Schw5}
where it was found numerically that several vortices can enter (or exit) at
once for disks with sufficiently large radius. In Sec.~III, where we
investigated the influence of $R_{1}/\xi $ (with $R_{2}/\xi =R/\xi =6.0$) we
only found $L\longrightarrow L+1$ transitions.

In order to show the complexity of the system under study, we investigated
one more field profile, the one shifted from the center of the disk i.e. a
ring magnetic field. We keep all parameters from the previous case, and
shift the field by $R_{d}=2.0\xi $ ($R_{1}-R_{d}=4.5\xi $, $%
R_{2}-R_{d}=6.0\xi $, $R/\xi =9.0$) towards the disk edge (a top view of
this profile is given schematically in the inset of Fig.~\ref{Figloop}). As
shown in Fig.~\ref{Figloop}, with $R_{d}\neq 0$ the multi-vortices become
more stable and for $H_{in}/H_{c2}>0.31$ are the ground state. The solid
curves represent the giant vortex states, dashed curves denote the energy of
the anti-vortex states while the dotted curves correspond to the energy of
the multi-vortices. Since giant vortex states show similar, {\it re-entrant}
behavior as for the $R_{d}=0$ case, the multi-vortex configurations behave
analogously. Giant-giant multi-vortex states dominate the free energy
diagram while the giant-ring combinations are present only as metastable
states and are not shown in Fig.~\ref{Figloop}. However, one should notice
the presence of giant--multi anti-vortex states, again strongly correlated
with {\it re-entrant} behavior. The complete equilibrium phase diagram for $%
R_{d}\in \lbrack 0.3,3]$ is given in the next section.

We checked that for the disk parameters which we used, an increase of the
number of components in Eq.~(\ref{SP}) does not lead to different vortex
configurations in the ground state. In order to investigate the region of
stability of multi-vortex states in the above analysis we took the order
parameter as a linear combination of three components in Eq.~(\ref{SP}) and
minimized the free energy with respect to the three variational parameters $%
C_{L_{i}}$. The giant and multi-vortex states considered before correspond
to the extremum points of the $F(C_{L_{1}},C_{L_{2}},C_{L_{3}})$ function.
This analysis has shown that accounting of third component in Eq.~(\ref{SP})
gives all states obtained in two component consideration. Additionally, it
results in 1) possible reducing of the region of existence of metastable
multi-vortex states at the low magnetic field limit, 2) appearance of
additional unstable states corresponding to the sadle points of the $%
F(C_{L_{1}},C_{L_{2}},C_{L_{3}})$ function, and, 3) no new vortex
configurations in the ground state. We did a similar investigation using a
numerical approach of Schweigert and Peeters \cite{Schw3}. This was done for
five component vortex configurations, and same results were obtained. Using
two different approaches, for three and five components, no new ground state
configurations are found. Moreover, energies of same states found in both
analysis for different number of components differ less than $0.2\%$.
Because all multi-vortex configurations considered in this paper are in the
ground state, or near it, we conclude that two components in Eq.~(\ref{SP})
are enough to describe the vortex structure in our system.

\section{$H_{in}-R$ PHASE DIAGRAMS}

First, we investigate the influence of the width of the positive magnetic
field region on the different vortex configurations. Having the free
energies of the different giant vortex configurations for several values of $%
R_{1}/\xi $, we construct an equilibrium vortex phase diagram. Fig.~\ref
{Fig13} shows this phase diagram for a superconducting disk with radius $%
R=6.0\xi $ and thickness $d=0.1\xi $ where we took $R_{2}=R$. The dashed
curves indicate where the ground state of the free energy changes from one $%
L $ state to another and the solid curve gives the normal/superconducting
transition. Notice that the superconducting/normal transition moves towards
lower fields with increasing radius of the positive field region. Also
notice that we don't have any negative $L$ (anti-vortex) state as ground
state even when the negative field area is much larger than the positive one.

As shown in previous section, an increase of $R_{1}/\xi $ is able to bring
the energy of the multi-vortex states below those of the giant vortex states
and they can become the ground state (see Fig.~\ref{Fig8}(a)). In our phase
diagram, the area bounded by the thick curves denotes the region of
existence of the giant-giant multi-vortex states as ground state. One can
see that, with enlargement of the positive field region, ground state
multi-vortices appear for $R_{1}/\xi =4.42$ and $H_{in}/H_{c2}=0.91$.
Further increase of $R_{1}/\xi $ broadens this {\it multi-vortex area} to
almost the whole superconducting region. These multi-vortex states consist
of vortices in a ring structure, and the total vorticity equals the highest
vorticity of the giant vortex states involved (at least in the case when no
anti-vortices are involved).

In order to show that the stabilization of the multi-vortex states due to an
inhomogeneous magnetic field is not peculiar to the $R=6.0\xi $ disks, we
repeated the previous calculations for a larger superconducting disk.
Furthermore, we investigated in detail the phenomena shown in previous
section (Figs.~\ref{Fig9}, \ref{Fig11}, \ref{Fig12}).

The effect of the size of the superconducting disk on the phase diagram is
illustrated in Fig.~\ref{Fig14}. The parameters considered are $R_{d}/\xi
=0.0$, $R_{1}/\xi =4.5$, and $R_{2}/\xi =6.0$. The solid lines indicate
where the ground state of the free energy changes from one state to another
(either giant or multi-vortex state) and dashed lines correspond to
transitions between different giant vortex states as metastable states. One
can clearly see the {\it re-entrant} behavior. For example, for $R/\xi =9.0$
we observe the change of the total vorticity as $L=0\rightarrow 1\rightarrow
2\rightarrow 3\rightarrow 1\rightarrow 0$. The $L=0$ and $L=1$ states as
ground state cover the largest part of the phase diagram. With increasing
disk size, all other giant vortex states are strongly suppressed in favor of
the various multi-vortex states. Different giant vortex states are present
as the ground state for small disk sizes but with increase of disk size,
islands with different multi-vortex configurations ({\it giant-ring} and 
{\it giant-giant} multi-vortex states) dominate the ground state diagram.
Precisely, for $R=6.38\xi $, giant-ring multi-vortex state appear as the
ground state, and for $R=6.77\xi $ we obtain the giant-giant multi-vortex
states as the ground state. In the previous diagram in Fig.~\ref{Fig13} we
have shown the existence of the giant-giant multi-vortex states as the
ground state for $R=6.0\xi $. With slight increase of the disk size, these
states become metastable, and with further enlargement of the
superconducting disk, they become the ground state again. However, these
states are different from the previous ones and in the contour plots of the
Cooper pair density of these ``new'' giant-giant multi-vortex states (see
Fig.~\ref{Fig12}) we found a giant vortex state in the middle of the
superconducting disk surrounded by anti-vortices, and where the {\it total}
vorticity is now equal to the lowest vorticity of the giant vortex states
that the multi-vortex consists of. For $R>12.33\xi $, the multi-vortex
states become metastable again and the $L=0$ and $L=1$ states become the
only ground states. Moreover, for $R>19.34\xi $ we have the Meissner state
for all values of the applied magnetic field.

To present the complexity and excitement in this study, we give one more
diagram. As shown before, moving the step-like field profile along the
radius of the superconducting disk, i.e. a ring magnetic field profile as in
the case of a current loop, stabilizes the giant-giant multi-vortex states
as the ground state. In Fig.~\ref{Fig15}, we present the phase diagram as
function of $R_{d}$. The parameters of the magnetic field profile are $%
R_{1}-R_{d}=4.5\xi $, $R_{2}-R_{d}=6.0\xi $, and $R/\xi =9.0$. Thick solid
curves indicate the transitions between different vortex states and dashed
lines denote the transitions between metastable giant vortex states. The 
{\it re-entrant} behavior, as in previous case, is clearly visible. The most
important result is that the multi-vortex states dominate this diagram.
Moreover, we have both types of multi-vortices, i.e. {\it giant-giant} and 
{\it giant-ring}, as stable and ground state, and with shifting the field
profile towards the disk periphery, giant-giant multi-vortex configurations
cover most of the superconducting region.. The giant-giant multi-vortex
states in this case appear both as states with no anti-vortices present, and
as giant--multi anti-vortex states. Following the transition lines, a
correspondence between different giant--multi anti-vortex states ($L_{1}$:$%
L_{2}$) can be seen: $L_{2}$ remains the same, while $L_{1}$ increases from $%
0$ to $2$. As one can see, the latter is strongly correlated with re-entrant
behavior. In the rest of the diagram, the multi-vortex states with the
``classical'' geometry (both {\it giant-giant} and {\it giant-ring})
dominate. However, a difference between those two states exists. As shown in
Fig.~\ref{Fig11}, in the case of a giant-ring state, vortices are
preferentially distributed within a ring-shaped lower density area. Further,
considering the high density areas at the disk periphery, a shift in phase
of $\Delta \theta =\pi /L$ is observed, in comparison with corresponding
giant-giant state (see, for example, Fig.~\ref{Fig_ent}), where $L$ is the
total vorticity. Although these states have a different origin, sometimes
they can exhibit a similar distribution of vortices, but the phase is always
able to distinguish between them.

\section{COMPARISON WITH REAL MAGNETIC FIELD PROFILE}

As emphasized before, our step-like field model is a simplification of the
magnetic field profile of a ferromagnetic dot (see solid curve in Fig.~\ref
{fig1}) as shown schematically in Fig.~\ref{Fig16}. In this case, the radius
of the dot corresponds to $R_{1}$, and $R_{2}$ is always equal to $R$. The
magnetic field and vector potential were calculated through numerical
integration of 
\begin{equation}
H_{z}(\rho ,z)=m_{z}\int\limits_{0}^{R_{1}}dr^{\prime }\frac{4r^{\prime
}(z^{\prime }-z)}{\sqrt{(r^{\prime }-\rho )^{2}+(z^{\prime }-z)^{2}}%
((r^{\prime }+\rho )^{2}+(z^{\prime }-z)^{2})}\left. E\left[ -\frac{%
4r^{\prime }\rho }{(r^{\prime }+\rho )^{2}+(z^{\prime }-z)^{2}}\right]
\right| _{z_{0}^{\prime }}^{z_{1}^{\prime }},
\end{equation}
and 
\[
A_{\varphi }(\rho ,z)=\frac{1}{\rho }\int\limits_{0}^{\rho }rH_{z}(r,z)dr, 
\]
where $m_{z}$ denotes the magnetic moment of a dot directed along the $z$%
-axis, $R_{1}$ is the radius of the dot, $z_{0}^{\prime }$ gives the
distance from the plane of interest, and $d_{d}=z_{1}^{\prime
}-z_{0}^{\prime }$ is the thickness of the dot. $E(x)$ is the complete
elliptic integral of the second kind. The calculations were done for $%
z\rightarrow 0$, which allowed us to neglect the radial component of the
magnetic field.

In order to show the quality of our previous model we repeated the analysis
from previous sections. First, we enlarge $R_{1}$, i.e. the radius of the
magnetic dot, investigating the influence of the magnetic field on the
vortex structure and especially on the stability of the multi-vortex states,
and secondly we studied the vortex configurations resulting from an increase
of the superconducting disk size.

For small values of $R_{1}/\xi $ the multi-vortex states are always
metastable states. However, with enlarging this parameter these states lower
in energy. The energies of the equilibrium vortex states as function of
magnetic moment of the dot measured in units of $m_{0}=H_{c2}$, are plotted
in Fig.~\ref{Fig17}(a) for radius of the dot $R_{1}/\xi =4.0$ and $R/\xi
=6.0 $ and in Fig.~\ref{Fig17}(b) for $R_{1}/\xi =4.0$ with a larger disk
radius $R/\xi =9.0$. Dashed curves correspond to anti-vortex states and
dotted curves represent the energy of the multi-vortex states. The
giant-giant multi-vortex states are given by $\left( L_{1}:L_{2}\right) $,
i.e. the angular momentum values they are composed of, and the giant-ring
multi-vortex states with lowest energy are given in Fig.~\ref{Fig17}(b) by $%
\left( (n_{1},L_{1}):\left( n_{2},L_{2}\right) \right) $. It should be noted
that there are many other metastable combinations possible, which are not
shown in the figures. Notice from Fig.~\ref{Fig17}(a) that the results are
qualitative similar to the results obtained with our step magnetic field
model (see Fig.~\ref{Fig8}). The difference is caused by the fact that part
of the negative magnetic flux does not penetrate the superconducting disk.
Because the positive field region in the center dictates the behavior of the
phase diagram even when the total flux is zero (see, for example, Fig.~\ref
{Fig8}(a)), we obtain very similar results with the real magnetic field
profile, but with an increased number of possible superconducting states.
However, increasing the size of the superconductor brings some qualitative
changes. First, it is clear from Fig.~\ref{Fig17}(b) that there is no {\it %
re-entrant} behavior, and, second, no giant-giant multi-vortex
configurations are ground state. But, this was expected since in our model
field profile the whole magnetic flux is trapped in the center of the disk,
with total flux equal zero. For the real profile, the total flux would be
zero if our disk is infinitely extended, and, more importantly, the flux is
now spread over the whole disk area. Naturally, smaller magnetic dots in
combination with large superconductors would make the correspondence better,
since most of the flux would be captured inside the disk.

A similar discussion holds for the results for the real magnetic field
profile of a current loop placed on top of a superconductor. This field
profile (solid curve) is compared with our model in Fig.~\ref{fig_lpr}
(thick dashed curve). The magnetic field and vector potential were
calculated numerically from 
\begin{equation}
H_{z}(\rho ,z)=\frac{Ik}{4\sqrt{R_{1}\rho }}\left[ \frac{R_{1}^{2}-\rho
^{2}-z^{2}}{(R_{1}-\rho )^{2}+z^{2}}E(k^{2})+K(k^{2})\right] \text{,}
\end{equation}
and

\begin{equation}
A_{\varphi }(\rho ,z)=\frac{I}{k}\sqrt{\frac{R_{1}}{\rho }}\left[ \left( 1-%
\frac{k^{2}}{2}\right) K(k^{2})-E(k^{2})\right] \text{,}
\end{equation}
with

\begin{equation}
k=2\sqrt{\frac{R_{1}\rho }{(R_{1}+\rho )^{2}+z^{2}}},
\end{equation}
where $K(x)$ is the complete elliptic integral of the first kind and $R_{1}$
denotes the radius of the loop with current $I$.

The free energy is shown in Fig.~\ref{Fig18} for a loop with radius $%
R_{1}=7.5\xi $, with superconductor disk size $R=9.0\xi $, as a function of
current $I$ measured in units of $I_{0}=\pi \xi H_{c2}/\mu _{0}$. Typical
values of $I_{0}$ are $3.29mA$, for aluminum, to $0.823A$ for
high-temperature superconductors. One should compare these results with
those presented in the $H_{in}-R_{d}$ phase diagram (Fig.~\ref{Fig15}) and
we find that our previous model contained all the essential physics of the
system. The free energy diagrams of both Fig.~\ref{Figloop} and Fig.~\ref
{Fig18} show the {\it re-entrant} behavior in total vorticity and the
existence of the giant-giant multi-vortex configurations over a large region
of the phase diagram.

\section{SUPERCONDUCTING DISK WITH A MAGNETIC DOT OR CURRENT LOOP ON TOP OF
IT IN THE PRESENCE OF A BACKGROUND HOMOGENEOUS EXTERNAL MAGNETIC FIELD}

In addition, we investigated the vortex structure of a superconducting disk
in the presence of an inhomogeneous magnetic field profile resulting from a
ferromagnetic dot and a homogeneous external background magnetic field. In
Figs.~\ref{Fig19}(a-c) we present the free energy as function of the
external homogeneous field, for different magnetic dot thicknesses. The
parameters were - radius of the dot $R_{1}/\xi =4.0$, radius of the
superconducting disk $R/\xi =6.0$ with thickness $d=0.1\xi $ and fixed
ferromagnetic dot profile which is shown as inset of Figs.~\ref{Fig19}(a-c).
As expected, when a negative external field overwhelms the average of the
positive magnetic field of the dot in the disk center, the anti-vortex
states become energetically more favorable, and, opposite, when the positive
external field becomes larger than the average negative value of the
ferromagnetic dot field, we see that the ground state goes through
successive giant vortex states towards the normal state. For example, for a
thickness of the magnetic dot $d_{d}=0.1\xi $ we obtain an almost
symmetrical figure with respect to $H_{ext}=-0.14H_{c2}$. However, with
increasing magnetic dot thickness, the magnetic field of the ferromagnetic
dot (given as insets in Figs.~\ref{Fig19}(a-c)) becomes more pronounced and
we obtain two sets of curves corresponding with the vortex and anti-vortex
states, each of which having two minima. These local minima occur at an
external field value which is approximately equal to the average field of
the positive or negative region of the magnetic dot profile. The region
between these two minima is characterized by a strong interplay of states.

For a current loop on top of the superconducting disk in a homogeneous
external field we obtained qualitatively similar results. Also two sets of
curves, with two local minima are visible in the free energy, for
sufficiently large currents in the loop (see Figs.~\ref{Fig20}(b-c)). We
show numerical results for a system consisting of a superconducting disk
with $R/\xi =9.0$ and thickness $d=0.1\xi $, and current loop with radius $%
R_{1}=7.5\xi $ (see Fig.~\ref{fig_lpr}). Just like in previous case, minima
in the free energy are related to average fields in positive and negative
region of the current loop magnetic field profile. However, several
differences exist with previous case. First, with increase of the thickness
of the magnetic dot in previous situation,\ the magnetic field profile
becomes more similar to our step-like field model. For the current loop case
this is not so, and with increasing current in the loop the magnetic field
becomes strongly inhomogeneous, rapidly increasing in the vicinity of the
loop. As a consequence the minimum in the free energy for negative external
field is lower in energy, and, moreover, the other minimum slowly diminishes
with further increase of the current in the loop. In this case even for
negative applied external field, the positive peak in the magnetic field
profile of the current loop is responsible for the stabilization of the
positive $L$-states, up to large $H_{ext}<0$ (see Fig.~\ref{Fig20}(d)). It
is obvious that the interplay between the external and the current loop
magnetic field exhibits more interesting physics due to the strong field
inhomogeneity.

\section{CONCLUSION}

We studied the superconducting state of a thin superconducting disk in the
presence of an inhomogeneous magnetic field. A ``model'' step-like magnetic
field profile was considered which is an approximation for the magnetic
field profile resulting from a magnetic disk or from a current loop. The
superconducting disk is assumed sufficiently thin that the magnetic field
produced by the superconducting currents can be neglected. The effect of the
width of the positive field region, the position of the field and the size
of the superconducting disk on the vortex configuration was investigated.
Numerous phase transitions were found, between states with different angular
momentum number and between giant and multi-vortex states, plus transitions
of the ground state between different multi-vortex configurations ({\it %
giant-ring} and {\it giant-giant} multi-vortex states). The model step-like
magnetic field (total flux through the system equals zero) is found to
stabilize the multi-vortex states both as the ground state (i.e., with
minimal energy) and as metastable states. Increase of the width of the
positive field region enhances the stability of the giant-giant
multi-vortices, while enlarging the superconducting disk decreases the
energy of the giant-ring multi-vortex state. In this case, the giant-giant
multi-vortex states can also be the ground state but it represents a new
configuration - a giant vortex surrounded by anti-vortices, and the {\it %
total} vorticity equals the lowest vorticity of the giant vortex states
which are involved. We found that with an increase of the disk size, {\it %
re-entrant} behavior of the total vorticity is possible. This behavior can
also be seen by shifting the field profile towards the disk boundary, i.e.
for the current loop magnetic field, when giant-giant multi-vortex
configurations cover most of the superconducting phase diagram region.
Comparison with results obtained for a real magnetic field profile shows
very good correspondence with our step-like models for the cases when most
of the magnetic flux is captured inside the superconducting disk. For larger
radius of magnetic dot, the positive magnetic field region of the
inhomogeneous magnetic field profile dominates and with increasing the
magnetic dot radius, the total flux in the superconductor also increases,
which is different from our model where the total flux always equals zero.
Thus, the positive part of the magnetic field profile determines mainly the
superconducting states, which leads to slightly different physics - no {\it %
re-entrant} behavior is present, more different superconducting states are
possible, but without the giant-giant multi-vortex states with
anti-vortices, since they are strongly correlated with a re-entrance of the
vorticity.

Addition of the background homogeneous external magnetic field, besides the
inhomogeneous field resulting from the magnetic dot, brings a qualitative
difference in the free energy diagram. For values of the external magnetic
field between the average field of the dot in the positive and the negative
region, we observe a strong interplay of the different superconducting giant
states. Moreover, with increase of the thickness of the dot, we obtain two
strong minima in the free energy, when the total field is approximately
equal to zero in the $\rho <R_{1}$, and the $\rho >R_{1}$ region,
respectively, where $R_{1}$ is the radius of the dot.

Here, we must emphasize that our results are only valid in the limit of very
thin disks. It allowed us to separate the two GL equations and thus we
neglected the magnetic field created by the superconducting currents.

\bigskip

\begin{center}
{\large ACKNOWLEDGMENTS}

\bigskip
\end{center}

This work was supported by the Flemish Science Foundation (FWO-VI), the
Belgian Inter-University Attraction Poles (IUAP-IV), the ``Onderzoeksraad
van de Universiteit Antwerpen'' (GOA), and the ESF programme on ``Vortex
matter''.

\bigskip

\begin{center}
{\large APPENDIX: Eigenfunctions of the linearized first GL equation in the
presence of an inhomogeneous magnetic field}
\end{center}

\subsection{Disk magnetic field profile}

The vector potential distribution is determined by the piecewise function

\begin{equation}
A\left( \rho \right) =\left\{ 
\begin{array}{c}
H_{0}\rho /2,\quad \qquad \qquad \qquad \ \ \ \ \ 0\leq \rho \leq
R_{1},\quad \ \ \ \ (I) \\ 
-H_{out}\rho /2+H_{out}R_{2}^{2}/2\rho ,\text{ \ \ \ \ }R_{1}\leq \rho \leq
R_{2},\text{ \ \ \ \ }(II) \\ 
\text{ \ \ }0,\qquad \qquad \qquad \qquad \quad \ \ \ \ \ R_{2}\leq \rho
\leq R.\quad \ \ \ \ (III)
\end{array}
\right.  \eqnum{A.1}  \label{vecpot1}
\end{equation}
The eigenfunctions of Eq.~(\ref{lin}) are expressed in the following way: 
\begin{equation}
f_{L,n}\left( \rho \right) =\left\{ 
\begin{array}{c}
f^{I}\left( \rho \right) ,\qquad \qquad \qquad \qquad (I) \\ 
b_{1}f_{1}^{II}\left( \rho \right) +b_{2}f_{2}^{II}\left( \rho \right)
,\qquad (II) \\ 
d_{1}f_{1}^{III}\left( \rho \right) +d_{2}f_{2}^{III}\left( \rho \right)
,\qquad (III)
\end{array}
\right.  \eqnum{A.2}  \label{eigfun1}
\end{equation}
where 
\begin{eqnarray*}
f^{I}\left( \rho \right) &=&\left( H_{0}\rho ^{2}/2\right) ^{\left| L\right|
/2}\exp \left( -H_{0}\rho ^{2}/4\right) M\left( -\nu _{n,I}\left( \Lambda
\right) ,\left| L\right| +1,H_{0}\rho ^{2}/2\right) , \\
f_{1}^{II}\left( \rho \right) &=&\left( H_{out}\rho ^{2}/2\right) ^{\left|
L_{II}^{\ast }\right| /2}\exp \left( -H_{out}\rho ^{2}/4\right) M\left( -\nu
_{n,II}\left( \Lambda \right) ,\left| L_{II}^{\ast }\right| +1,H_{out}\rho
^{2}/2\right) , \\
f_{2}^{II}\left( \rho \right) &=&\left( H_{out}\rho ^{2}/2\right) ^{\left|
L_{II}^{\ast }\right| /2}\exp \left( -H_{out}\rho ^{2}/4\right) U\left( -\nu
_{n,II}\left( \Lambda \right) ,\left| L_{II}^{\ast }\right| +1,H_{out}\rho
^{2}/2\right) , \\
f_{1}^{III}\left( \rho \right) &=&J_{\left| L\right| }\left( \sqrt{1+\Lambda 
}\rho \right) , \\
f_{2}^{III}\left( \rho \right) &=&Y_{\left| L\right| }\left( \sqrt{1+\Lambda 
}\rho \right) ,
\end{eqnarray*}
with

\begin{eqnarray*}
L_{II}^{\ast } &=&L-H_{out}R_{2}^{2}/2, \\
\nu _{n,I}\left( \Lambda \right) &=&-\frac{1+\left| L\right| -L}{2}+\frac{%
1+\Lambda }{2H_{0}}, \\
\nu _{n,II}\left( \Lambda \right) &=&-\frac{1+\left| L_{II}^{\ast }\right|
+L_{II}^{\ast }}{2}+\frac{1+\Lambda }{2H_{out}},
\end{eqnarray*}
$J_{m}\left( x\right) $ and $Y_{m}\left( x\right) \ $are the Bessel
functions of the first and second kind, $M\left( a,c,y\right) $ and $U\left(
a,c,y\right) $\ are the Kummer functions. To find the unknown constants $%
b_{1\left( 2\right) }$, $d_{1\left( 2\right) }$ and the eigenvalue $\Lambda $
we have to join the different parts of $f_{L,n}\left( \rho \right) $ and its
derivatives at $R_{1}$ and $R_{2}$ as well as to use the boundary condition $%
\left. \left( \partial f/\partial \rho \right) \right| _{\rho =R}=0.$

To simplify the next calculation let us introduce the following notations: $%
K_{1}=f^{I}\left( R_{1}\right) $, $I_{1}=f_{1}^{II}\left( R_{1}\right) $, $%
N_{1}=f_{2}^{II}\left( R_{1}\right) $, $I_{3}=f_{1}^{II}\left( R_{2}\right) $%
, $N_{3}=f_{2}^{II}\left( R_{2}\right) $, $P_{3}=f_{1}^{III}\left(
R_{2}\right) $,\ $Q_{3}=f_{2}^{III}\left( R_{2}\right) $,\ $K_{2}=\left. 
\frac{\partial f^{I}\left( \rho \right) }{\partial \rho }\right| _{\rho
=R_{1}}$,\ $I_{2}=\left. \frac{\partial f_{1}^{II}\left( \rho \right) }{%
\partial \rho }\right| _{\rho =R_{1}}$,\ $N_{2}=\left. \frac{\partial
f_{2}^{II}\left( \rho \right) }{\partial \rho }\right| _{\rho =R_{1}}$, $%
I_{4}=\left. \frac{\partial f_{1}^{II}\left( \rho \right) }{\partial \rho }%
\right| _{\rho =R_{2}}$,\ $N_{4}=\left. \frac{\partial f_{2}^{II}\left( \rho
\right) }{\partial \rho }\right| _{\rho =R_{2}}$,\ $P_{4}=\left. \frac{%
\partial f_{1}^{III}\left( \rho \right) }{\partial \rho }\right| _{\rho
=R_{2}}$,\ $Q_{4}=\left. \frac{\partial f_{2}^{III}\left( \rho \right) }{%
\partial \rho }\right| _{\rho =R_{2}}$,\ $P_{5}=\left. \frac{\partial
f_{1}^{III}\left( \rho \right) }{\partial \rho }\right| _{\rho =R}$,\ $%
Q_{5}=\left. \frac{\partial f_{2}^{III}\left( \rho \right) }{\partial \rho }%
\right| _{\rho =R}.$ With these notations the system of equations which
determine the coefficients $b_{1\left( 2\right) }$, $d_{1\left( 2\right) }$
and $\Lambda $ have the simple form 
\begin{eqnarray}
b_{1}I_{1}+b_{2}N_{1} &=&K_{1},  \nonumber \\
b_{1}I_{2}+b_{2}N_{2} &=&K_{2},  \nonumber \\
b_{1}I_{3}+b_{2}N_{3}-d_{1}P_{3}-d_{2}Q_{3} &=&0,  \eqnum{A.3}
\label{sysconst1} \\
b_{1}I_{4}+b_{2}N_{4}-d_{1}P_{4}-d_{2}Q_{4} &=&0,  \nonumber \\
d_{1}P_{5}+d_{2}Q_{5} &=&0.  \nonumber
\end{eqnarray}
From the first four equations of the system~(\ref{sysconst1}) we obtain: 
\begin{eqnarray}
b_{1} &=&\frac{K_{1}N_{2}-K_{2}N_{1}}{I_{1}N_{2}-I_{2}N_{1}},  \eqnum{A.4}
\label{const1} \\
b_{2} &=&\frac{I_{1}K_{2}-I_{2}K_{1}}{I_{1}N_{2}-I_{2}N_{1}},  \nonumber
\end{eqnarray}
\[
d_{1}=\frac{\left( K_{1}N_{2}-K_{2}N_{1}\right) \left(
I_{3}Q_{4}-Q_{3}I_{4}\right) +\left( I_{1}K_{2}-I_{2}K_{1}\right) \left(
Q_{4}N_{3}-Q_{3}N_{4}\right) }{P_{3}Q_{4}-P_{4}Q_{3}}, 
\]
\[
d_{2}=\frac{\left( K_{1}N_{2}-K_{2}N_{1}\right) \left(
P_{3}I_{4}-I_{3}P_{4}\right) +\left( I_{1}K_{2}-I_{2}K_{1}\right) \left(
N_{4}P_{3}-N_{3}P_{4}\right) }{P_{3}Q_{4}-P_{4}Q_{3}}. 
\]
Inserting these results into the last equation of system~(\ref{sysconst1})
results into the non-linear equation for $\Lambda $: 
\begin{equation}
P_{5}\left[ \left( K_{1}N_{2}-K_{2}N_{1}\right) \left(
I_{3}Q_{4}-Q_{3}I_{4}\right) +\left( I_{1}K_{2}-I_{2}K_{1}\right) \left(
Q_{4}N_{3}-Q_{3}N_{4}\right) \right] \qquad  \eqnum{A.5}  \label{eqlam1}
\end{equation}
\[
+Q_{5}\left[ \left( K_{1}N_{2}-K_{2}N_{1}\right) \left(
P_{3}I_{4}-I_{3}P_{4}\right) +\left( I_{1}K_{2}-I_{2}K_{1}\right) \left(
N_{4}P_{3}-N_{3}P_{4}\right) \right] =0. 
\]
To obtain the correct $\Lambda $ values we have to exclude from the spectrum
of solutions of Eq.~(\ref{eqlam1}) those which results in zeros of both
denominators $I_{1}N_{2}-I_{2}N_{1}$ and $P_{3}Q_{4}-P_{4}Q_{3}$.

In the large radius limit $R\rightarrow \infty $, i.e. which is equivalent
to the thin film limit, the Bessel functions in the (III)-region have to be
replaced by their asymptotics. Substituting them in the equation $%
d_{1}P_{5}+d_{2}Q_{5}=0$\ we obtain 
\begin{equation}
\Lambda =-1+\frac{1}{R^{2}}\left[ \frac{\pi }{4}\left( 3+2|L|\right)
-\arctan \frac{d_{1}}{d_{2}}\right] ^{2}.  \eqnum{A.6}
\end{equation}
$f_{L,n}\left( \rho \right) $ near the edge of the sample is equal to $\sqrt{%
d_{1}^{2}+d_{2}^{2}}$.

\subsection{Ring magnetic field profile}

The vector potential distribution in this case is

\begin{equation}
A\left( \rho \right) =\left\{ 
\begin{array}{c}
0,\ \ \ \ \ \qquad \qquad \qquad \ \ \ \ \ \ \ \ \ \ \ 0\leq \rho \leq R_{d},%
\text{\ \ \ \ }(I) \\ 
H_{0}\rho /2-H_{0}R_{d}^{2}/2\rho ,\text{ \ \ \ \ \ \ \ \ \ \ }R_{d}\leq
\rho \leq R_{1},\text{ \ }(II) \\ 
-H_{out}\rho /2+H_{out}R_{2}^{2}/2\rho ,\text{ \ \ \ \ }R_{1}\leq \rho \leq
R_{2},\ \ (III) \\ 
\text{ \ }0,\ \ \ \ \qquad \qquad \qquad \ \ \ \ \ \ \ \ \ \ \ R_{2}\leq
\rho \leq R,\ \ \ \ (IV)
\end{array}
\right.  \eqnum{A.7}  \label{vecpot2}
\end{equation}
and, therefore, the eigenfunctions of Eq.~(\ref{lin}) are expressed as
follows:

\begin{equation}
f_{L,n}\left( \rho \right) =\left\{ 
\begin{array}{c}
f^{I}\left( \rho \right) ,\qquad \qquad \qquad \qquad (I) \\ 
b_{1}f_{1}^{II}\left( \rho \right) +b_{2}f_{2}^{II}\left( \rho \right) ,%
\text{ \ \ \ \ \ }(II) \\ 
d_{1}f_{1}^{III}\left( \rho \right) +d_{2}f_{2}^{III}\left( \rho \right) ,%
\text{ \ \ \ \ }(III) \\ 
e_{1}f_{1}^{IV}\left( \rho \right) +e_{2}f_{2}^{IV}\left( \rho \right)
,\qquad (IV)
\end{array}
\right.  \eqnum{A.8}  \label{eigfun2}
\end{equation}
where 
\begin{eqnarray*}
f^{I}\left( \rho \right) &=&f_{1}^{IV}\left( \rho \right) =J_{\left|
L\right| }\left( \sqrt{1+\Lambda }\rho \right) , \\
f_{1}^{II}\left( \rho \right) &=&\left( H_{0}\rho ^{2}/2\right) ^{\left|
L_{II}^{\ast }\right| /2}\exp \left( -H_{0}\rho ^{2}/4\right) M\left( -\nu
_{n,II}\left( \Lambda \right) ,\left| L_{II}^{\ast }\right| +1,H_{0}\rho
^{2}/2\right) , \\
f_{2}^{II}\left( \rho \right) &=&\left( H_{0}\rho ^{2}/2\right) ^{\left|
L_{II}^{\ast }\right| /2}\exp \left( -H_{0}\rho ^{2}/4\right) U\left( -\nu
_{n,II}\left( \Lambda \right) ,\left| L_{II}^{\ast }\right| +1,H_{0}\rho
^{2}/2\right) , \\
f_{1}^{III}\left( \rho \right) &=&\left( H_{out}\rho ^{2}/2\right) ^{\left|
L_{III}^{\ast }\right| /2}\exp \left( -H_{out}\rho ^{2}/4\right) M\left(
-\nu _{n,III}\left( \Lambda \right) ,\left| L_{III}^{\ast }\right|
+1,H_{out}\rho ^{2}/2\right) , \\
f_{2}^{III}\left( \rho \right) &=&\left( H_{out}\rho ^{2}/2\right) ^{\left|
L_{III}^{\ast }\right| /2}\exp \left( -H_{out}\rho ^{2}/4\right) U\left(
-\nu _{n,III}\left( \Lambda \right) ,\left| L_{III}^{\ast }\right|
+1,H_{out}\rho ^{2}/2\right) , \\
f_{2}^{IV}\left( \rho \right) &=&Y_{\left| L\right| }\left( \sqrt{1+\Lambda }%
\rho \right) ,
\end{eqnarray*}
with 
\begin{eqnarray*}
L_{II}^{\ast } &=&L+H_{0}R_{d}^{2}/2, \\
L_{III}^{\ast } &=&L-H_{out}R_{2}^{2}/2, \\
\nu _{n,II}\left( \Lambda \right) &=&-\frac{1+\left| L_{II}^{\ast }\right|
-L_{II}^{\ast }}{2}+\frac{1+\Lambda }{2H_{0}}, \\
\nu _{n,III}\left( \Lambda \right) &=&-\frac{1+\left| L_{III}^{\ast }\right|
+L_{III}^{\ast }}{2}+\frac{1+\Lambda }{2H_{out}}.
\end{eqnarray*}
The unknown constants $b_{1\left( 2\right) },d_{1\left( 2\right)
},e_{1\left( 2\right) }$ and the eigenvalue $\Lambda $ are found by joining
the different parts of $f_{L,n}\left( \rho \right) $ and its derivatives at $%
R_{d},R_{1}$ and $R_{2}$ as well as from the boundary condition $\left.
\left( \partial f/\partial \rho \right) \right| _{\rho =R}=0.$ The
corresponding system of equations has the form: 
\begin{eqnarray}
b_{1}I_{1}+b_{2}N_{1} &=&K_{1},  \nonumber \\
b_{1}I_{2}+b_{2}N_{2} &=&K_{2},  \nonumber \\
b_{1}I_{3}+b_{2}N_{3}-d_{1}P_{3}-d_{2}Q_{3} &=&0,  \nonumber \\
b_{1}I_{4}+b_{2}N_{4}-d_{1}P_{4}-d_{2}Q_{4} &=&0,  \eqnum{A.9}
\label{sysconst2} \\
d_{1}P_{5}+d_{2}Q_{5}-e_{1}S_{5}-e_{2}T_{5} &=&0,  \nonumber \\
d_{1}P_{6}+d_{2}Q_{6}-e_{1}S_{6}-e_{2}T_{6} &=&0,  \nonumber \\
e_{1}S_{7}+e_{2}T_{7} &=&0.  \nonumber
\end{eqnarray}
where the following notations were introduced:$K_{1}=f^{I}\left(
R_{d}\right) $,\ $I_{1}=f_{1}^{II}\left( R_{d}\right) $,\ $%
N_{1}=f_{2}^{II}\left( R_{d}\right) $,\ $K_{2}=\left. \frac{\partial
f^{I}\left( \rho \right) }{\partial \rho }\right| _{\rho =R_{d}}$, $%
I_{2}=\left. \frac{\partial f_{1}^{II}\left( \rho \right) }{\partial \rho }%
\right| _{\rho =R_{d}}$,\ $N_{2}=\left. \frac{\partial f_{2}^{II}\left( \rho
\right) }{\partial \rho }\right| _{\rho =R_{d}}$,\ $I_{3}=f_{1}^{II}\left(
R_{1}\right) $, $N_{3}=f_{2}^{II}\left( R_{1}\right) $,\ $%
P_{3}=f_{1}^{III}\left( R_{1}\right) $,\ $Q_{3}=f_{2}^{III}\left(
R_{1}\right) $,\ $I_{4}=\left. \frac{\partial f_{1}^{II}\left( \rho \right) 
}{\partial \rho }\right| _{\rho =R_{1}}$, $N_{4}=\left. \frac{\partial
f_{2}^{II}\left( \rho \right) }{\partial \rho }\right| _{\rho =R_{1}}$,\ $%
P_{4}=\left. \frac{\partial f_{1}^{III}\left( \rho \right) }{\partial \rho }%
\right| _{\rho =R_{1}}$, $Q_{4}=\left. \frac{\partial f_{2}^{III}\left( \rho
\right) }{\partial \rho }\right| _{\rho =R_{1}}$, $P_{5}=f_{1}^{III}\left(
R_{2}\right) $,\ $Q_{5}=f_{2}^{III}\left( R_{2}\right) $,\ $%
S_{5}=f_{1}^{IV}\left( R_{2}\right) $,\ $T_{5}=f_{2}^{IV}\left( R_{2}\right) 
$, $P_{6}=\left. \frac{\partial f_{1}^{III}\left( \rho \right) }{\partial
\rho }\right| _{\rho =R_{2}}$, $Q_{6}=\left. \frac{\partial
f_{2}^{III}\left( \rho \right) }{\partial \rho }\right| _{\rho =R_{2}}$, $%
S_{6}=\left. \frac{\partial f_{1}^{IV}\left( \rho \right) }{\partial \rho }%
\right| _{\rho =R_{2}}$, $T_{6}=\left. \frac{\partial f_{2}^{IV}\left( \rho
\right) }{\partial \rho }\right| _{\rho =R_{2}}$, $S_{7}=\left. \frac{%
\partial f_{1}^{IV}\left( \rho \right) }{\partial \rho }\right| _{\rho =R}$, 
$T_{7}=\left. \frac{\partial f_{2}^{IV}\left( \rho \right) }{\partial \rho }%
\right| _{\rho =R}$. The constants $b_{1\left( 2\right) }$ and $d_{1\left(
2\right) }$ are determined by the expressions~(\ref{const1}) and 
\[
e_{1}=\frac{\left( P_{5}T_{6}-T_{5}P_{6}\right) d_{1}+\left(
Q_{5}T_{6}-T_{5}Q_{6}\right) d_{2}}{S_{5}T_{6}-S_{6}T_{5}}, 
\]
\[
e_{2}=\frac{\left( S_{5}P_{6}-P_{5}S_{6}\right) d_{1}+\left(
S_{5}Q_{6}-Q_{5}S_{6}\right) d_{2}}{S_{5}T_{6}-S_{6}T_{5}}. 
\]
From the last equation of system~(\ref{sysconst2}) 
\begin{equation}
\begin{array}{l}
\left[ \left( K_{1}N_{2}-K_{2}N_{1}\right) \left(
I_{3}Q_{4}-Q_{3}I_{4}\right) +\left( I_{1}K_{2}-I_{2}K_{1}\right) \left(
Q_{4}N_{3}-Q_{3}N_{4}\right) \right] \\ 
\times \left[ S_{7}\left( P_{5}T_{6}-T_{5}P_{6}\right) +T_{7}\left(
S_{5}P_{6}-P_{5}S_{6}\right) \right] \\ 
+\left[ \left( K_{1}N_{2}-K_{2}N_{1}\right) \left(
P_{3}I_{4}-I_{3}P_{4}\right) +\left( I_{1}K_{2}-I_{2}K_{1}\right) \left(
N_{4}P_{3}-N_{3}P_{4}\right) \right] \\ 
\times \left[ S_{7}\left( Q_{5}T_{6}-T_{5}Q_{6}\right) +T_{7}\left(
S_{5}Q_{6}-Q_{5}S_{6}\right) \right] =0
\end{array}
\eqnum{A.10}
\end{equation}
we obtain the spectrum of $\Lambda $ values (notice, that we have to exclude
the zeros of the denominators $I_{1}N_{2}-I_{2}N_{1}$, $%
P_{3}Q_{4}-P_{4}Q_{3} $ and $S_{5}T_{6}-S_{6}T_{5}$).

In the large radius limit $R\rightarrow \infty $, similar considerations
have to be made as for the previous case of the disk profile. $f_{L,n}\left(
\rho \right) \ $near the sample edge is equal to $\sqrt{e_{1}^{2}+e_{2}^{2}}$%
\ and 
\begin{equation}
\Lambda =-1+\frac{1}{R^{2}}\left[ \frac{\pi }{4}\left( 3+2|L|\right)
-\arctan \frac{e_{1}}{e_{2}}\right] ^{2}.  \eqnum{A.11}
\end{equation}

\begin{figure}[tbp]
\caption{The magnetic field profile of the magnetic disk (solid curve) and
the corresponding model step profile (dashed curve).}
\label{fig1}
\end{figure}
\begin{figure}[tbp]
\caption{The magnetic field dependence of the lowest eigenvalues of the
linearized first GL equation for different angular momenta $L$ and for (a)$%
~R_{1}/\protect\xi =1.5$, (b) $R_{1}/\protect\xi =3.0$, and (c)$~R_{1}/%
\protect\xi =4.5$. The corresponding curves for the first radial excited
state, i.e. $n=1$, are presented in figures (d-f). The top axis gives the
flux through the positive magnetic field region in units of the flux
quantum. }
\label{fig2}
\end{figure}
\begin{figure}[tbp]
\caption{The Cooper pair density for the giant vortex states with angular
momenta $L=0,1,2$ (solid curves), the antivortex states $L=-1,-2$ (dotted
curves) and the ring vortex state $n=1$, $L=0$ (dashed curve) at the
magnetic field $H_{in}=0.75H_{C2}$ for (a)$~R_{1}/\protect\xi =1.5$, (b) $%
R_{1}/\protect\xi =3.0$, and (c)$~R_{1}/\protect\xi =4.5$.}
\label{fig_psi}
\end{figure}
\begin{figure}[tbp]
\caption{The free energy of the giant vortex states with different angular
momenta $L$ as a function of the applied magnetic field in the positive
region for (a)$~R_{1}/\protect\xi =1.5$, (b) $R_{1}/\protect\xi =3.0$, and
(c)$~R_{1}/\protect\xi =4.5$. Only the states with $\left| L\right| \leq 10$
are shown. The insets depict the magnetic field dependence of the disk
magnetization for the ground giant vortex state for different values of the $%
R_{1}/\protect\xi $ parameter. Figure (d) gives the positive flux captured
in the superconducting disk for the different $L$-ground states as the
ground state as function of $R_{1}/\protect\xi $. }
\label{Fig3}
\end{figure}
\begin{figure}[tbp]
\caption{The Cooper pair density for the giant vortex states with angular
momenta $L=0,1,2$ (solid curves), the anti-vortex states $L=-1,-2$ (dotted
curves) and the ring-shaped vortex state $n=1$, $L=0$ (dashed curve) at the
magnetic field $H_{in}=0.75H_{C2}$ for (a)$~R/\protect\xi =9.0$, and (b) $R/%
\protect\xi =12.0$.}
\label{fig_psi1}
\end{figure}
\begin{figure}[tbp]
\caption{The Cooper pair density for (a) the Meissner state ($L=0$), and (b)
the ring-shaped vortex state $n=1$, $L=0$ at the magnetic field $%
H_{in}=0.75H_{c2}$ for different sizes of the superconducting disk.}
\label{fig_psi2}
\end{figure}
\begin{figure}[tbp]
\caption{The free energy of the giant vortex states with different angular
momenta $L$ as a function of the applied magnetic field for (a)$~R/\protect%
\xi =9.0$, (b) $R/\protect\xi =12.0$, (c)$~R/\protect\xi =18.0$, and (d) $R/%
\protect\xi =20.0$, with $R_{1}/\protect\xi =4.5$, and $R_{2}/\protect\xi
=6.0$. Dashed curves represent the energy of the anti-vortex states. }
\label{Fig5}
\end{figure}
\begin{figure}[tbp]
\caption{The magnetic field dependence of the disk magnetization for the
ground giant vortex state corresponding to the states in Fig.~\ref{Fig5},
for (a) $R/\protect\xi =9.0$, (b) $R/\protect\xi =12.0$, and (c)$~R/\protect%
\xi =18.0$. }
\label{Fig6}
\end{figure}
\begin{figure}[tbp]
\caption{The free energy of the giant vortex states with different angular
momenta $L$ as a function of the external magnetic field for (a)$~R_{1}/%
\protect\xi =5.25$, $R_{2}/\protect\xi =R/\protect\xi =6.0$, and (b) $R_{1}/%
\protect\xi =4.5$, $R_{2}/\protect\xi =6.0$, $R/\protect\xi =9.0$. Dashed
curves illustrate the energy of the anti-vortex states. In (a) dotted curves
represent the giant-giant multivortex states while in (b) dotted curves
depict the free energy of the giant-ring multi-vortex states, and the solid
curves correspond to the giant-giant states. Thick solid curve illustrates
the energy of the ring vortex state, i.e. $(n,L)=(1,0)$}
\label{Fig8}
\end{figure}
\begin{figure}[tbp]
\caption{Transition between two giant vortex states shown through the
contour plots of the superconducting wave function density for the ($0:4$)
giant-giant multi-vortex state for different values of the magnetic field in
the positive field region (corresponding to the magnetization of the dot).}
\label{Fig_ent}
\end{figure}
\begin{figure}[tbp]
\caption{Contour plots of the superconducting density for the ground state
and the corresponding phase contour plots (see Fig.~\ref{Fig8}(b)) for
different values of the magnetic field in the positive region.}
\label{Fig9}
\end{figure}
\begin{figure}[tbp]
\caption{Contourplot of the superconducting density for the $((5,0)$:$(0,1))$
giant-ring multi-vortex state (see Fig.~\ref{Fig8}(b)) for different values
of the applied magnetic field.}
\label{Fig11}
\end{figure}
\begin{figure}[tbp]
\caption{(a-c) Contour plots of the superconducting density for the $(0$:$7)$%
, $\left( 0\text{:}8\right) $, and, $\left( 0\text{:}9\right) $ giant-multi
anti-vortex states, respectively, for different values of the magnetic field 
$H_{in}$; (d-f) the corresponding contour plots of the phase of the
superconducting wave function density. Notice that the phase near the
boundary is near $0$ or $2\protect\pi $ but due to the finite numerical
accuracy it oscillates between $2\protect\pi -\protect\varepsilon $ and $2%
\protect\pi +\protect\varepsilon $ where $\protect\xi \sim 10^{-5}$. }
\label{Fig12}
\end{figure}
\begin{figure}[tbp]
\caption{The free energy of the giant vortex states with different angular
momenta $L$ as a function of the external magnetic field for $R_{d}=2.1%
\protect\xi $, $R_{1}=6.6\protect\xi $, $R_{2}=8.1\protect\xi $, $R/\protect%
\xi =9.0$. Dashed curves depict the anti-vortex states and dotted curves
represent the free energy of the giant-giant multi-vortex states. A bird
view of the magnetic field profile is given in the inset. }
\label{Figloop}
\end{figure}
\begin{figure}[tbp]
\caption{The $H_{in}-R_{1}$ equilibrium vortex phase diagram for a thin
superconducting disk with $R_{d}/\protect\xi =0.0$, $R_{2}/\protect\xi =6.0$%
, and $R/\protect\xi =6.0$. Dashed curves indicate transitions between
different giant vortex states and the thick shaded area denotes the
multi-vortex region. The normal/superconducting state transition is given by
the thin solid curve.}
\label{Fig13}
\end{figure}
\begin{figure}[tbp]
\caption{The $H_{in}-R$ phase diagram for the ground state of a thin
superconducting disk with $R_{d}/\protect\xi =0.0$, $R_{1}/\protect\xi =4.5$%
, and $R_{2}/\protect\xi =6.0$. Solid curves indicate transitions between
different vortex states including multi-vortex (giant-multi anti-vortex and
ring-giant) regions. Dashed curves denote the transitions between different
metastable giant vortex states.}
\label{Fig14}
\end{figure}
\begin{figure}[tbp]
\caption{The $H_{in}-R_{d}$ phase diagram for the ground state of a thin
superconducting disk with $R_{1}-R_{d}=4.5\protect\xi $ and $R_{2}-R_{d}=6.0%
\protect\xi $, i.e. a ring inhomogeneous magnetic field distribution. The
same curve convention is used as in Fig. \ref{Fig14}. }
\label{Fig15}
\end{figure}
\begin{figure}[tbp]
\caption{The configuration: a superconducting disk with radius $R$ and
thickness $d$ with a ferromagnetic dot with radius $R_{1}$ and thickness $%
d_{d}$, which is placed on top of it.}
\label{Fig16}
\end{figure}
\begin{figure}[tbp]
\caption{The free energy of different vortex states as a function of the
magnetic moment of the ferromagnetic dot with (a)$~R_{1}/\protect\xi =4.5$, $%
R_{2}/\protect\xi =R/\protect\xi =6.0$, and (b) $R_{1}/\protect\xi =4.5$, $%
R_{2}/\protect\xi =R/\protect\xi =9.0$. In (a), dotted curves depict the
free energy of the giant-giant multi-vortex states while in (b) dotted
curves denote giant-ring states. The bold curve gives the energy of the ring
vortex state, i.e. $(n,L)=(1,0)$. The energy of the anti-vortex states is
given by the dashed curves.}
\label{Fig17}
\end{figure}
\begin{figure}[tbp]
\caption{The magnetic field profile as produced by a current loop (solid
curve) and its corresponding model profile (dashed curve).}
\label{fig_lpr}
\end{figure}
\begin{figure}[tbp]
\caption{The free energy of different vortex states as a function of the
current in a loop with (a)$~R_{1}/\protect\xi =7.5$, $R_{2}/\protect\xi =R/%
\protect\xi =9.0$. Dashed curves depict the free energy of the anti-vortex
states and dotted curves denote the giant-giant multi-vortex states. }
\label{Fig18}
\end{figure}
\begin{figure}[tbp]
\caption{The free energy of giant vortex states as a function of the
external magnetic field in a superconducting disk with magnetic dot on top
of it with parameters $R_{1}/\protect\xi =4.0$, $R/\protect\xi =6.0$ and (a) 
$d_{d}=0.1\protect\xi $, (b) $d_{d}=0.5\protect\xi $, and (c) $d_{d}=1.0%
\protect\xi $. Dashed curves depict the free energy of the anti-vortex
states. Insets show the magnetic field profile inside the superconducting
disk created by the magnetic dot. }
\label{Fig19}
\end{figure}
\begin{figure}[tbp]
\caption{The free energy of giant vortex states as a function of the
external magnetic field in a superconducting disk with current loop on top
of it with parameters $R_{1}/\protect\xi =7.5$, $R/\protect\xi =9.0$ and (a) 
$I/I_{0}=5.0$, (b) $I/I_{0}=10.0$, (c) $I/I_{0}=15.0$, and (d) $I/I_{0}=20.0$%
. Dashed curves depict the free energy of the anti-vortex states. Insets
show the magnetic field profile inside the superconducting disk created by
the current loop.}
\label{Fig20}
\end{figure}

\end{document}